\documentclass[english,a4paper,manuscript]{revtex4}
\usepackage[T1]{fontenc}
\usepackage[latin1]{inputenc}

\makeatletter

\makeatletter

\makeatletter

\usepackage{geometry}

\makeatother

\makeatother

\usepackage{babel}
\makeatother
\begin{document}

\title{On the nature of the so-called generic instabilities in dissipative
relativistic hydrodynamics}

\author{A. L. Garcia-Perciante$^{1}$, L. S. Garcia-Colin$^{2}$, A. Sandoval-Villalbazo$^{3}$}

\address{$^{1}$Depto. de Matematicas Aplicadas y Sistemas, Universidad Autonoma
Metropolitana-Cuajimalpa, Artificios 40 Mexico D.F 01120, Mexico.\\
 $^{2}$Depto. de Fisica, Universidad Autonoma Metropolitana-Iztapalapa,
Av. Purisima y Michoacan S/N, Mexico D. F. 09340, Mexico. Also at
El Colegio Nacional, Luis Gonzalez Obregon 23, Centro Historico, Mexico
D. F. 06020, Mexico. \\
 $^{3}$Depto. de Fisica y Matematicas, Universidad Iberoamericana,
Prolongacion Paseo de la Reforma 880, Mexico D. F. 01210, Mexico}

\begin{abstract}
It is shown that the so-called generic instabilities that appear in
the framework of relativistic linear irreversible thermodynamics,
describing the fluctuations of a simple fluid close to equilibrium,
arise due to the coupling of heat with hydrodynamic acceleration which
appears in Eckart's formalism of relativistic irreversible thermodynamics.
Further, we emphasize that such behavior should be interpreted as
a contradiction to the postulates of linear irreversible thermodynamics
(LIT), namely a violation of Onsager's hypothesis on the regression
of fluctuations, and not as fluid instabilities. Such contradictions
can be avoided within a relativistic linear framework if a Meixner-like
approach to the phenomenological equations is employed. 
\end{abstract}
\maketitle

\section{Introduction}

Relativistic irreversible thermodynamics has had a rather peculiar
history. The first two proposals on the subject came from C. Eckart
\cite{Eckart} in 1940 and from L. Landau and I. Lifshitz in the early
fifties \cite{Landau}. Since both of them have been shown to be particular
cases of the so-called 'first order theories' \cite{HL1}, we concentrate
here on the former proposal, mainly based on the construction of an
energy stress tensor where heat flux is included, thus incorporating
heat \emph{in the same status as a mechanical energy}. Moreover it
is also built so to satisfy the first law of thermodynamics. It was
later shown by W. Israel and coworkers \cite{Israel,Israel & Stewart}
that both theories may lead to transport equations of the parabolic
type, therefore violating causality. Also, it has been recently suggested
that heat, being a non-mechanical form of energy, should not be included
in the stress-energy tensor to be coupled to Einstein's tensor in
the context of general relativity \cite{mexmeeting07}. Regarding
causality, we will leave additional comments on this interpretation
and appropriate status in statistical physics for a future publication.
Here we want to concentrate on the role of heat in deriving transport
equations.

Indeed, besides these difficulties, during the last two decades of
the past century, several papers appeared pointing out another shortcoming
of the two theories mentioned earlier namely, that they predict hydrodynamic
instabilities in the linear regime \cite{Landau,HL1,Israel & Stewart}.
This interpretation is not correct. What they really show is that
fluctuations under such conditions do not obey the Onsager's regression
assumption. We remind the reader that this hypothesis states that
spontaneous fluctuations that occur in a fluid in equilibrium evolve
towards the equilibrium state following linear equations for the macroscopic
variables \cite{ONSG}. This fact led workers in the field to consider
the so-called 'second order theories' \cite{Israel,HL2} as those
which correctly solve the two main objections. This approach has now
been taken as the basis of a new version of a modified form for the
relativistic hydrodynamic equations suitable to deal with the hot
dense matter produced in the Relativistic Heavy Ion Collider, RHIC
\cite{RHIC}.

In this work we show that the problems mentioned above are caused
by the structure of the constitutive equation for the heat flux proposed
by Eckart which couples it with the hydrodynamic acceleration. This
relation, when introduced as a closure to the system of hydrodyamic
equations, yields the so-called generic instabilities. It is important
to point out that this coupling, whose thermodynamic meaning has always
been questioned, has so far not been supported by experiment \cite{oviedo}.

As an alternative, we study the effect of a Fourier-like constitutive
equation as can be obtained, for example, from Meixner's formulation
of linear irreversible thermodynamics \cite{meix,degroot}. This theory
is based upon the idea that the heat flux is a component of a separate
total energy flux \cite{LGCSJNET06} and leads to a system of linearized
transport equations which correctly describe the evolution of the
fluctuations according to LIT. Moreover, the second law of thermodynamics
is neatly derived and the stress-energy tensor maintains its canonical
form including only state variables.

These results suggest that those attempts seeking to deal with these
problems basing their ideas on Extended Irreversible Thermodynamics,
EIT \cite{EIT}, may be unnecessary. The results obtained from A Meixner-like
approach are free from additional parameters other than the standard
transport coefficients. On the other hand, EIT, whose constitutive
equations are of the Maxwell-Cattaneo type introduces additional parameters
that, on the long run, have to be adjusted. Other difficulties with
EIT theories have been extensively discussed in the literature \cite{LGCSMOLPHYS}.

To pursue our discussion we have structured the paper as follows.
In section II the relativistic transport equations are derived from
Eckart's proposal for the stress-energy tensor. Section III contains
the linearized system and the consequent derivation of the so-called
instability of the transverse velocity mode. Final remarks, including
the alternative proposal corresponding to a Meixner-like scheme, are
included in the last section of this work.

\section{Relativistic transport equations: Eckart's formalism}

The evolution of a relativistic fluid is described by the balance
equations considered together with suitable constitutive relations.
The continuity equation in relativistic hydrodynamics is a statement
of conservation of particles. For a non-reactive fluid and denoting
the particle flux as $N^{\mu}=nu^{\mu}$, such an equation reads

\noindent \begin{equation}
N_{;\mu}^{\mu}=0\label{eq:1}\end{equation}
 where $n$ is the particle number density. Here and throughout this
work greek indices run from 1 to 4, latin indices from 1 to 3, and
a semicolon indicates a covariant derivative. The fluid four-velocity
is $u^{\mu}$, so Eq. (\ref{eq:1}) can be written as\begin{equation}
\dot{n}+n\theta=0\label{eq:2}\end{equation}
 where $\theta=u_{;\mu}^{\mu}$ and a dot denotes a total time derivative.

The general conservation equation for the stress-energy tensor\begin{equation}
T_{\nu;\mu}^{\mu}=0\label{eq:3}\end{equation}
 encompasses both energy and momentum balances. In Eckart's formalism,
$T^{\mu\nu}$ is given by\begin{equation}
T_{\nu}^{\mu}=\frac{n\varepsilon}{c^{2}}u^{\mu}u_{\nu}+ph_{\nu}^{\mu}+\pi_{\nu}^{\mu}+\frac{1}{c^{2}}q^{\mu}u_{\nu}+\frac{1}{c^{2}}u^{\mu}q_{\nu}\label{eq:4}\end{equation}
 where $\varepsilon$ is the energy per particle in the comoving frame,
$p$ is the local pressure, $\pi_{\nu}^{\mu}$ is the Navier tensor,
$q^{\mu}$ is the heat flux and $h_{\nu}^{\mu}$ is a spatial projector
defined, for a $(+++-)$ signature, as\begin{equation}
h_{\nu}^{\mu}=\delta_{\nu}^{\mu}+\frac{u^{\mu}u_{\nu}}{c^{2}}\label{eq:5}\end{equation}
 The last two terms in Eq. (\ref{eq:4}) are identified as the heat
flux contributions included in Eckart's formalism as part of the proposed
relativistic generalization. Also, these terms satisfy orthogonality
conditions equivalent to those for the viscous dissipation namely,
\begin{equation}
u_{\mu}\pi_{\nu}^{\mu}=u^{\nu}\pi_{\nu}^{\mu}=0,\qquad q_{\mu}u^{\mu}=q^{\mu}u_{\mu}=0\label{eq:5a}\end{equation}
 It is worthwhile at this point to emphasize that these heat flux
terms are not present in the stress tensor in the non-relativistic
theory as well as in the relativistic version of Meixner's thermodynamics
\cite{LGCSJNET06}.

Equation (\ref{eq:3}), after computing the covariant derivative of
$T_{\nu}^{\mu}$ given by Eq. (\ref{eq:4}), using Eq. (\ref{eq:2})
and the operator $\dot{(\,)}=u^{\alpha}(\,)_{;\alpha}$ yields the
momentum balance equation

\begin{eqnarray}
\left(\frac{n\varepsilon}{c^{2}}+\frac{p}{c^{2}}\right)\dot{u}_{\nu}+\left(\frac{n\dot{\varepsilon}}{c^{2}}+\frac{p}{c^{2}}\theta\right)u_{\nu}+p_{,\mu}h_{\nu}^{\mu}+\pi_{\nu;\mu}^{\mu}\nonumber \\
+\frac{1}{c^{2}}\left(q_{;\mu}^{\mu}u_{\nu}+q^{\mu}u_{\nu;\mu}+\theta q_{\nu}+u^{\mu}q_{\nu;\mu}\right) & = & 0\label{eq:6}\end{eqnarray}
 The evolution of the internal energy is given by the projection of
Eq. (\ref{eq:3}):\begin{equation}
u^{\nu}T_{\nu;\mu}^{\mu}=0\label{eq:7}\end{equation}
 from which one can obtain the equation\begin{equation}
n\dot{\varepsilon}+p\theta+u_{,\mu}^{\nu}\pi_{\nu}^{\mu}+q_{;\mu}^{\mu}+\frac{1}{c^{2}}\dot{u}^{\nu}q_{\nu}=0\label{eq:8}\end{equation}
 where use has been made of the fact that, from Eq. (\ref{eq:5a}),
$u^{\nu}q_{\nu;\mu}=-u_{;\mu}^{\nu}q_{\nu}$ and a similar relation
holds for the viscous term. It is convenient to recast this equation
in terms of the temperature as state variable instead of $\varepsilon$
by means of the local equilibrium assumption. That is, since $\varepsilon=\varepsilon\left(n,\, T\right)$,
one can write\begin{equation}
\dot{\varepsilon}=\left(\frac{\partial\varepsilon}{\partial n}\right)_{T}\dot{n}+\left(\frac{\partial\varepsilon}{\partial T}\right)_{n}\dot{T}\label{eq:9}\end{equation}
 Using the relations $\left(\frac{\partial\varepsilon}{\partial n}\right)_{T}=-\frac{T\beta}{n^{2}\kappa_{T}}+\frac{p}{n^{2}}$
and $\left(\frac{\partial\varepsilon}{\partial T}\right)_{n}=C_{n}$
where $\beta$ is the volume expansion coefficient, $\kappa_{T}$
the isothermal compressibility and $C_{n}$ the specific heat at constant
$n$, Eq. (\ref{eq:8}) can be written as\begin{equation}
nC_{n}\dot{T}+\left(\frac{T\beta}{\kappa_{T}}\right)\theta+u_{;\mu}^{\nu}\pi_{\nu}^{\mu}+q_{;\mu}^{\mu}+\frac{1}{c^{2}}\dot{u}^{\nu}q_{\nu}=0\label{eq:10}\end{equation}

Equations (\ref{eq:2}), (\ref{eq:6}) and (\ref{eq:10}), the conservation
equations, form an incomplete set. In order to obtain the dynamics
of the state variables one has to introduce constitutive relations
as closure equations. In order to propose these phenomenological expressions
for the fluxes $\pi^{\mu\nu}$ and $q^{\mu}$ , one calculates the
entropy production from the entropy balance equation which is obtained
by means of the local equilibrium assumption, that is, if the entropy
per particle is $s=s\left(n,\,\varepsilon\right)$,\begin{equation}
\dot{s}=\left(\frac{\partial s}{\partial n}\right)_{\varepsilon}\dot{n}+\left(\frac{\partial s}{\partial\varepsilon}\right)_{n}\dot{\varepsilon}\label{eq:11}\end{equation}
 Substitution of Eqs. (\ref{eq:2}) and (\ref{eq:8}) in Eq. (\ref{eq:11})
and using the thermostatic relations \begin{equation}
\left(\frac{\partial s}{\partial n}\right)_{\varepsilon}=-\frac{p}{n^{2}T}\,,\qquad\left(\frac{\partial s}{\partial\varepsilon}\right)_{n}=\frac{1}{T}\,,\label{eq:12}\end{equation}
 yields an entropy density balance equation which can be brought to
the following general structure\begin{equation}
n\dot{s}+J_{[s];\nu}^{\nu}=\sigma\label{eq:13}\end{equation}
 where $J_{[s]}^{\nu}$ is identified as an entropy density flux and
the entropy production is given by the expression\begin{equation}
\sigma=-\frac{q^{\nu}}{T}\left(\frac{T_{,\nu}}{T}+\frac{T}{c^{2}}\dot{u}_{\nu}\right)-\frac{u_{,\mu}^{\nu}}{T}\pi_{\nu}^{\mu}\label{eq:14}\end{equation}
 In order to assure the positiveness of $\sigma$, and thus satisfy
the second law of thermodynamics, Eckart proposes the following constitutive
relations\begin{equation}
\pi_{\nu}^{\mu}=-2\eta h_{\alpha}^{\mu}h_{\nu}^{\beta}\tau_{\beta}^{\alpha}-\zeta\theta\delta_{\nu}^{\mu}\label{eq:c}\end{equation}
 \begin{equation}
q^{\nu}=-\kappa h_{\mu}^{\nu}\left(T^{,\mu}+\frac{T}{c^{2}}\dot{u}^{\mu}\right)\label{eq:17}\end{equation}
 where the transport coefficients involved are the bulk viscosity
$\zeta$, the shear viscosity $\eta$ and the thermal conductivity
$\kappa$. In Eq. (\ref{eq:c}) $\tau_{\nu}^{\mu}$ is the traceless
symmetric part of the velocity gradient tensor. Equation (\ref{eq:17})
deserves a closer look. In it, the first term in parenthesis corresponds
to the usual Fourier-type constitutive equation. The second term,
which arises from the inclusion of the heat terms in the stress tensor,
is not in the canonical form namely, it cannot be considered a thermodynamic
force. This objection is equally applicable to the second term of
Eq. (\ref{eq:14}). A complete discussion of this issue can be found
in Ref. \cite{mexmeeting07} and will not be repeated here. However
this fact is brought to the attention of the reader since, as will
be shown in the next section, this term leads to the so-called instability
found by Hiscock and Lindblom \cite{HL1}.

\section{Linearized equations FOR THE FLUCTUATIONS}

In this section, we shall study the behavior of the linearized equations
which result from Eckart's scheme when spontaneous fluctuations of
the state variables occur around equilibrium. These equations arise
upon substitution of the constitutive equations given by Eqs. (\ref{eq:c}-\ref{eq:17})
into the general conservation equations (\ref{eq:2}), (\ref{eq:6})
and (\ref{eq:10}). Next we linearize by setting $T=T_{0}+\delta T$,
$u^{k}=\delta u^{k}$ ($\delta u^{4}=0$) and $\theta=\delta\theta$,
which according to Eq. (\ref{eq:2}) is equivalent to the condition
$n=n_{0}+\delta n$. Here the naught subscripts characterize their
equilibrium values and $\delta$ denotes the corresponding fluctuation.
The ensuing process requires a minimum effort. In fact, by simple
inspection, one notices that the second bracket and the first three
terms in the third bracket in Eq. (\ref{eq:6}) contain at least quadratic
terms in the fluctuations as well as the third and fifth terms in
Eq. (\ref{eq:10}). Therefore, the only terms that introduce a difference
with the results obtained in ordinary relativistic hydrodynamics are
$u^{4}q_{\nu;4}$ in the momentum balance and $q_{;\mu}^{\mu}$ in
the energy balance equations. Both arise from the inclusion of heat
in the stress tensor.

The rest of the procedure is straightforward. Using the well known
techniques of standard linearized non-relativistic hydrodynamics \cite{mountain,berne,boon}
we first write the linearized equations,

\begin{equation}
\delta\dot{n}+n_{0}\delta\theta=0\label{eq:18}\end{equation}
 \begin{eqnarray}
\frac{1}{c^{2}}\left(n_{0}\varepsilon_{0}+p_{0}\right)\delta\dot{u}_{\nu}+\frac{1}{\kappa_{T}}\delta n_{,\nu}+\frac{1}{\beta\kappa_{T}}\delta T_{,\nu}\nonumber \\
-\zeta\delta\theta_{,\nu}-2\eta\left(\delta\tau_{;\nu}^{\mu}\right)_{;\mu}-\frac{\kappa}{c^{2}}\delta\dot{T}_{,\nu}-\frac{\kappa T_{0}}{c^{4}}\delta\ddot{u}_{\nu} & =0\label{eq:19}\end{eqnarray}
 \begin{equation}
nC_{n}\delta\dot{T}+\left(\frac{T_{0}\beta}{\kappa_{T}}\right)\delta\theta-\kappa\left(\delta T^{,k}+\frac{T_{0}}{c^{2}}\delta\dot{u}^{k}\right)_{;k}=0\label{eq:20}\end{equation}
 where use has been made of the fact that $p=p\left(n,\, T\right)$.

Equations (\ref{eq:18})-(\ref{eq:20}) are the linearized equations
for thermodynamic fluctuations in Eckart's version of relativistic
hydrodynamics. The next step is to separate $\delta\theta$ from the
transverse velocity. For this purpose we calculate the curl and the
divergence of Eq. (\ref{eq:19}). The second operation yields\begin{eqnarray}
-\frac{\kappa T_{0}}{c^{4}}\delta\ddot{\theta}+\frac{1}{c^{2}}\left(n_{0}\varepsilon_{0}+p_{0}\right)\delta\dot{\theta}+\frac{1}{\kappa_{T}}\nabla^{2}\delta n+\frac{1}{\beta\kappa_{T}}\nabla^{2}\delta T\nonumber \\
-\left(\zeta+\frac{4}{3}\eta\right)\nabla^{2}\delta\theta-\frac{\kappa}{c^{2}}\nabla^{2}\delta\dot{T} & = & 0\label{eq:22}\end{eqnarray}
 This is a scalar differential equation with unknowns $\delta\theta$
(or $\delta n$) and $\delta T$ . We shall come back to it afterwards.
On the other hand, the first operation, recalling that the curl of
gradient vanishes, yields\begin{equation}
\frac{\kappa T_{0}}{c^{4}}\delta\ddot{U}_{\nu}-\frac{1}{c^{2}}\left(n_{0}\varepsilon_{0}+p_{0}\right)\delta\dot{U}_{\nu}+2\eta\nabla^{2}\delta U_{\nu}=0\label{eq:23}\end{equation}
 where $\delta U_{\alpha}=\epsilon_{\alpha\nu}^{\mu}{\delta u}_{;\mu}^{\nu}$
are the components of the curl of the velocity field. $\epsilon_{\alpha\nu}^{\mu}$
is the well known Levi-Civita's tensor. Thus, \emph{this procedure
decouples the transverse mode fluctuations from the system of hydrodynamic
equations}, yielding an independent equation. Calculating a Fourier-Laplace
transform with transform variables $k^{\ell}$ and $s$ respectively,
we obtain\begin{equation}
\delta\hat{\tilde{U}}_{\nu}\left(k^{\ell},\, s\right)=\frac{\delta\hat{U}_{\nu}\left(k^{\ell},\,0\right)\left[\frac{\kappa T_{0}}{c^{4}}s-\frac{1}{c^{2}}\left(n_{0}\varepsilon_{0}+p_{0}\right)\right]+\frac{\kappa T_{0}}{c^{4}}\left.\frac{\partial\delta\hat{U}_{\nu}\left(k^{\ell},\, t\right)}{\partial t}\right|_{t=0}}{\frac{\kappa T_{0}}{c^{4}}s^{2}-\frac{1}{c^{2}}\left(n_{0}\varepsilon_{0}+p_{0}\right)s-2\eta k^{2}}\label{eq:235}\end{equation}
 where $\delta\hat{\tilde{U}}_{\nu}$ is the Laplace-Fourier transform
of $\delta U_{\nu}$. Thus, the exponential growth, or decay, of the
transverse component of the velocitiy fluctuations will depend on
the roots of the denominator, namely, the solution of the dispersion
relation\begin{equation}
\frac{\kappa T_{0}}{c^{4}}s^{2}-\left(\frac{n_{0}\varepsilon_{0}+p_{0}}{c^{2}}\right)s-2\eta k^{2}=0\label{eq:24}\end{equation}
 Equation (\ref{eq:24}) has two real roots, namely\begin{equation}
s=\frac{c^{2}}{2\kappa T_{0}}\left[n_{0}\varepsilon_{0}+p_{0}\pm\sqrt{\left(n_{0}\varepsilon_{0}+p_{0}\right)^{2}+8k^{2}\eta\kappa T_{0}}\right]\label{eq:25}\end{equation}
 which is precisely the result the authors of Ref. \cite{HL1} arrived
at in a more laborious way. Notice that, not only the procedure here
developed yields the dispersion relation in a more clear, direct way
but it also highlights two important points. Firstly, the equation
for the transverse mode is completely decoupled from the system. Thus,
it is unnecessary to make any assumptions about the direction of the
velocity fluctuations as done in that work. The second and most important
point we wish to emphasize is the fact that, as a consequence of decoupling
of both components of the velocity field, longitudinal and transverse,
one can easily identify the source of the exponential growth found
in the transverse mode fluctuations. Indeed, for spatially homogeneous
perturbations, $k=0$, Eq. (\ref{eq:25}) yields a positive root\begin{equation}
s=\frac{n_{0}\varepsilon_{0}+p_{0}}{\kappa T_{0}}c^{2}\label{eq:26}\end{equation}
 Equation (\ref{eq:26}) is a critical result. It clearly indicates
that Onsager's assumption of the regression of fluctuations is violated.
This is in open contradiction with the tenets of LIT. In this sense,
Eckarts's formalism is suspect.

\section{Discussion}

The results obtained in the previous section, have motivated the formulation
and use of generalized constitutive equations. However, as can be
clearly seen by inspection of Eqs. (\ref{eq:22}-\ref{eq:24}) its
cause is precisely the presence of the acceleration term in the constitutive
equation for the heat flux, Eq. (\ref{eq:17}) which, in turn can
be traced down to the phenomenological formalism first introduced
by Eckart. The coupling proposed in Eq. (\ref{eq:17}) is thus responsible
for the theory being cathegorized as unphysical, and thus displaced
by others \cite{GL}.

On the other hand, an alternative proposal for the stress energy tensor,
the one consistent with Meixner's theory can be easily shown to yield
a system of hydrodynamic equations in which fluctuations behave canonically
this instability is absent. Indeed, as discussed in Ref. \cite{LGCSJNET06},
the stress energy tensor there assumed is\begin{equation}
T_{\mu}^{\nu}=\rho u^{\nu}u_{\mu}+ph_{\mu}^{\nu}+\pi_{\mu}^{\nu}\label{eq:27}\end{equation}
 Equation (\ref{eq:27}) provides an alternative manner to introduce
heat without identifying it as a state variable, consistently with
classical thermodynamics. Heat is a form of energy, energy in transit,
and must be included in a total energy balance equation. In Meixner's
irreversible thermodynamics, heat flux is firstly introduced in a
total energy balance given by\begin{equation}
J_{[E];\nu}^{\nu}=0\label{eq:275}\end{equation}
 where $J_{[E]}^{\nu}$ is the total energy flux and includes the
mechanic and internal energy fluxes as well as the dissipative heat
flux\begin{equation}
J_{[E]}^{\nu}=\rho c^{2}u^{\mu}+n\varepsilon u^{\nu}+q^{\nu}\,.\label{eq:28}\end{equation}
 This procedure yields an entropy production in terms of products
of thermodynamic forces and fluxes exclusively,\begin{equation}
\sigma=-q^{\nu}\frac{T_{,\nu}}{T^{2}}-\frac{u_{;\nu}^{\mu}}{T}\pi_{\mu}^{\nu}\label{eq:29}\end{equation}
 which is consistent with Clausius' idea of uncompensated heat and
motivates a constitutive equation for the heat flux where the acceleration
term is absent. Indeed, the constitutive relations in this formalism
correspond to laws of proportionality between forces and fluxes which
ultimately results in a momentum balance equation where the second
derivative of the velocity is absent, and thus the evolution equation
for the fluctuation of the velocity variable $\delta U_{\nu}$ reads\begin{equation}
\left(\frac{n_{0}\varepsilon_{0}+p_{0}}{c^{2}}\right)\delta\dot{U}_{\nu}=2\eta\nabla^{2}\delta U_{\nu}\label{eq:30}\end{equation}
 which, performing a similar analysis as the one that lead to Eq.
(\ref{eq:235}) yields the dispersion relation\begin{equation}
-\left(\frac{n_{0}\varepsilon_{0}+p_{0}}{c^{2}}\right)s-2\eta k^{2}=0\label{eq:31}\end{equation}
 Equation (\ref{eq:31}) clearly predicts only an exponential decay
in time for $\delta U_{\nu}$. This result is compared with the unphysical
behavior found in the previous section by analyzing the non-relativistic
limit of both results for $k\neq0$ in Appendix A. As shown in this
appendix, the root which generates the exponential growth mentioned
above contains the thermal conductivity \emph{even in the non-relativistic
limit}. This is completely at odds with classical hydrodynamics and
with the fact that the heat terms in Eckart's tensor, Eq. (\ref{eq:4}),
are assumed to be strictly relativistic.

To finish this discussion, we would like to go back to Eqs. (\ref{eq:20})
and (\ref{eq:22}), a set of coupled equations for $\delta T$ and
$\delta\theta$ (or $\delta n$). To solve this system one has to
go to the frequency and wave number representation and transform them
into a set of two coupled algebraic equations. With the solutions
obtained one can calculate the density-density or temperature-temperature
correlation functions. The former one, as well known \cite{mountain,berne,boon},
is related to the so-called dynamic structure factor whose form, for
a simple fluid, is known as the Rayleigh-Brillouin spectrum. It consists
of a central, or Rayleigh peak and two symmetrically located peaks
known as the Brillouin peaks. Rayleigh's peak has a width in frequency
$\Delta\omega\propto D_{th}k^{2}$ where $D_{th}=\frac{\kappa}{\rho C_{v}}$
is the thermal diffusivity and $k=2\pi\lambda^{-1}$ when calculated
with Eq. (\ref{eq:27}). On the other hand, if Eckart's tensor is
used, it appears the the Rayleigh-Brillouin expectrum does not exist.
This fact would be in open contradiction with experiment. Calculations
along this line will be published elsewhere.

\appendix

\section{Non-relativistic limit}

In this appendix, the non-relativistic limit of Eqs. (\ref{eq:24})
and (\ref{eq:31}) are explored. In such limit, one can identify $n_{0}\varepsilon_{0}\rightarrow\rho_{0}c^{2}$
and further notice that $p_{0}\ll\rho_{0}c^{2}$ where $\rho_{0}$
is the equilibrium mass density. Introducing these facts in the dispersion
relation obtained within Eckart's formalism, Eq. (\ref{eq:24}), yields\begin{equation}
\frac{\kappa T_{0}}{c^{4}}s^{2}-\rho_{0}s-2\eta k^{2}=0\label{eq:33}\end{equation}
 while Eq. (\ref{eq:31}) yields the linear equation\begin{equation}
-\rho_{0}s-2\eta k^{2}=0\label{eq:34}\end{equation}
 Equation (\ref{eq:33}) has two roots which, using a binomial expansion
for $8\kappa T_{0}\eta k^{2}\ll\rho^{2}c^{4}$, are\begin{equation}
s_{1}\simeq-\frac{2\eta k^{2}}{\rho}\label{eq:35}\end{equation}
 \begin{equation}
s_{2}\simeq\frac{\rho_{0}c^{4}}{\kappa T_{o}}+\frac{2\eta k^{2}}{\rho_{0}}\label{eq:36}\end{equation}
 Notice that the first root, $s_{1}$, corresponds to the decaying
behavior found within Meixner's approach, Eq. (\ref{eq:34}). This
behavior, in which the velocity fluctuations decay due to viscous
effects is more physical than the one given by the second root $s_{2}$,
which corresponds to the so-called instability referred to in Ref.
\cite{HL1}. This behavior of the velocity fluctuations is due to
both viscous and thermal effects, a result which we insist, is untenable
in this limit.

\end{document}